\documentclass[pdflatex,sn-mathphys-num]{sn-jnl}

\usepackage{graphicx}%
\usepackage{multirow}%
\usepackage{amsmath,amssymb,amsfonts}%
\usepackage{amsthm}%
\usepackage{mathrsfs}%
\usepackage[title]{appendix}%
\usepackage{xcolor}%
\usepackage{textcomp}%
\usepackage{manyfoot}%
\usepackage{booktabs}%
\usepackage{algorithm}%
\usepackage{algorithmicx}%
\usepackage{algpseudocode}%
\usepackage{listings}%
\usepackage{comment}
\usepackage{tabularx}
\usepackage{array}

\theoremstyle{thmstyleone}%
\newtheorem{theorem}{Theorem}
\theoremstyle{thmstyletwo}%
\newtheorem{remark}{Remark}%

\theoremstyle{thmstylethree}%

\newcommand{\R}{\mathbb{R}}

\newtheorem{lemma}{Lemma}
\newtheorem{corollary}{Corollary}

\begin{document}

\title[Beyond the Gap]{Beyond the Gap: What Imaginary-Time Evolution Reveals About the Edge of the Continuum}


\author*[1]{\fnm{Ruben} \sur{Karapetyan}} 

\author[1]{\fnm{Aleš} \sur{Wodecki}}

\author[2]{\fnm{Denys} \sur{Bondar}}

\author[1]{\fnm{Jakub} \sur{Mareček}}

\affil*[1]{ \orgname{Czech Technical University in Prague}}

\affil[2]{\orgname{
Tulane University}}


\abstract{The spectral gap, the energy separating the ground state from the first excitation, governs many quantum phenomena, yet it is far harder to compute than the ground state. We propose a mathematically rigorous method for extracting the gap from imaginary-time evolution in infinite-dimensional Hamiltonians, where the first excitation typically marks the edge of a continuous spectrum. We show that a gap is not automatically detected: detection depends on how the initial state and observable couple the ground state to excitations at the bottom of the continuum. When this coupling follows a power law, the gap is recovered together with a logarithmic correction whose coefficient yields the exponent of the spectral density near the edge of the continuum, which reflects the physics of the excitations. We confirm both predictions numerically for an exactly solvable model. Our results offer a new numerical approach to many-body quantum physics, in which both the gap and the threshold exponent are obtained merely at the cost of a ground-state calculation.}




\maketitle

\section{Introduction}\label{sec1}

The spectral gap of a Hamiltonian $H$, the distance between its ground-state energy and the next point of its spectrum, governs a remarkable range of quantum phenomena. However, for general Hamiltonians, even the question whether the gap is positive is undecidable \cite{Cubitt2022Undecidability}. In practice, determining a gap requires information about the spectrum above the ground state, and while a large arsenal of methods exists for approximating ground states, the corresponding tools for excited states are far less developed. Methods that recover the gap from a ground-state computation alone, reusing this existing machinery, are therefore of considerable interest.

Imaginary-time evolution is precisely such a ground-state computation. Evolving an initial state $\psi$ under $e^{-\tau H}$ and renormalizing projects it exponentially fast onto the ground state, and this is the workhorse behind a wealth of numerical methods, from quantum Monte Carlo to tensor-network algorithms. The key observation of Lukin et al.~\cite{Lukin2024} is that the \emph{rate} at which this projection happens equals the spectral gap. In particular, for a suitable initial state and a self-adjoint observable $O$, the expectation of the commutator $[H,O]$ along the normalized imaginary-time trajectory $\phi(\tau)$ decays as 
\begin{equation}\label{EqLukin}
    \ln\left|\langle\phi(\tau),[H,O]\phi(\tau)\rangle\right| = -\Delta\tau+O(1), \qquad \tau\to\infty,
\end{equation}
where $\Delta$ is the gap between the ground state and the first excited energy. This approach led to a lightweight algorithm that made it possible to compute numerically, for the first time, the spectral gap of the three-dimensional transverse-field Ising model in the thermodynamic limit~\cite{Lukin2024}. Bilokon et al.~\cite{Bilokon2026} extended it to another important physical quantity: dispersion relations.

There is, however, a gap in the theory behind the spectral gap. The proof of Eq.~\eqref{EqLukin} was carried out only for finite-dimensional Hamiltonians, and it was left as a conjecture in Ref.~\cite{Lukin2024} that the method should also apply when the first excited energy sits at the threshold of a continuous spectrum. This is exactly the situation one meets in the thermodynamic limit, in scattering problems, and in essentially every infinite-dimensional model of physical interest, where the spectrum is no longer a collection of eigenvalues but a continuum.

In this paper we address this question by placing the method on a rigorous spectral-theoretic footing for general self-adjoint Hamiltonians on infinite-dimensional Hilbert spaces. The answer turns out to be more subtle, and more interesting, than Eq.~\eqref{EqLukin} alone suggests.

Our contributions are as follows. In Theorem~\ref{thm:main_2_gaps}, we first treat Hamiltonians whose ground-state and first-excited energies are both isolated eigenvalues; here the analysis closely parallels the finite-dimensional case of \cite{Bondar2023}, and the first gap is recovered exactly as in Eq.~\eqref{EqLukin}. The genuinely new phenomena appear when only the ground state is isolated and the first excited energy is the bottom of a continuum. Theorem~\ref{thm:main_1_gap} shows that the commutator expectation is then controlled by a Laplace-type integral $F(\tau)$ against a signed spectral measure $\nu_H$ that couples the ground state to the excited spectral subspace through the observable, and that the gap is visible in the asymptotics if and only if $F(\tau)$ decays subexponentially. Crucially, the mere existence of a gap, together with a nontrivial overlap between the initial state and the excited subspace, is \emph{not} enough: the coupling measure may fail to charge a neighborhood of the threshold, or its imaginary part may cancel, and in either case the method is blind to the gap. In Theorem~\ref{thm:polynomial_density}, we then give a clean sufficient condition in terms of the threshold behavior of $\nu_H$: if its density behaves like $c\,x^{\alpha}$ with $\alpha>-1$ near the first excitation threshold, the gap is recovered, and the decay acquires an additional logarithmic correction, 
\begin{equation}\label{EqRuben}
    \ln\left|\langle\phi(\tau),[H,O]\phi(\tau)\rangle\right| = -\Delta\tau-(\alpha+1)\ln\tau+O(1).
\end{equation}
which has no counterpart in Eq.~\eqref{EqLukin}. Note that for infinite-dimensional systems a purely linear fit to extract $\Delta$ is systematically biased, and the correct ansatz must include a $\ln\tau$. Since the coefficient of the $\ln\tau$ term in Eq.~\eqref{EqRuben} is $-(\alpha+1)$, fitting this ansatz to the computed commutator expectation allows us to reconstruct computationally the threshold exponent $\alpha$ of the spectral density. We then verify the threshold condition explicitly for a classical solvable model, 
 the three-dimensional one-center point interaction in Section~\ref{sec:point_interaction}, where the exponent is $\alpha=\tfrac12$
and we present a numerical illustration in Section~\ref{sec:numerics} 
that confirms both the predicted gap and the predicted logarithmic correction.

\section{Spectral theorem for unbounded operators}
\label{sec:spectral_theorem}

By the Spectral Theorem \cite{Conway1990,Kato1995,Kowalski2009,ReedSimon1980,Rudin1991,Teschl2009}, every self-adjoint (bounded or unbounded) operator \(H\) on \(\mathcal H\)
determines a unique projection-valued measure
\[
P:\mathcal B(\mathbb R)\to \mathcal L(\mathcal H).
\]
Equivalently, \(P\) determines a resolution of the identity
\[
P_\lambda := P((-\infty,\lambda]), \qquad \lambda\in\mathbb R.
\]
With respect to this resolution, \(H\) admits the representation
\[
H=\int_{\mathbb R} \lambda\,dP_\lambda,
\]
where the equality is understood on the domain
\[
D(H)=
\left\{
\psi\in\mathcal H:
\int_{\mathbb R}\lambda^2\,d\langle P_\lambda\psi,\psi\rangle<\infty
\right\}.
\]
Moreover, for every bounded Borel function \(f\), the functional calculus is
given by
\[
f(H)=\int_{\mathbb R} f(\lambda)\,dP_\lambda \in \mathscr B(\mathcal{H}).
\]
Assuming that
\[
\sigma(H)\subset \{\lambda_0,\lambda_1\}\cup[\lambda_2,\infty), \quad 
-\infty<\lambda_0 <\lambda_1<\lambda_2,
\quad \lambda_0,\lambda_1\in\sigma(H),
\]
is equivalent to assuming
\[
H=\int_{\mathbb R}\lambda\,dP_\lambda
=
\lambda_0 P(\{\lambda_0\})
+
\lambda_1  P(\{\lambda_1\})
+
\int_{[\lambda_2,\infty)} \lambda\,dP_\lambda .
\]
Analogously, assuming that
\[
\sigma(H)\subset \{\lambda_0\}\cup[\lambda_1,\infty), \quad \lambda_0 <\lambda_1,
\quad \lambda_0,\lambda_1\in\sigma(H),
\]
is equivalent to assuming
\[
H=\int_{\mathbb R}\lambda\,dP_\lambda
=
\lambda_0 P(\{\lambda_0\})
+
\int_{[\lambda_1,\infty)} \lambda\,dP_\lambda .
\]

We can directly define complex-valued spectral measures from the projection-valued measures as in \cite{Conway1990,Kato1995,Kowalski2009,ReedSimon1980,Rudin1991,Teschl2009}.
To a Hamiltonian $H$ and two states $\varphi, \phi$ from the Domain of $H$, we assign a spectral complex-valued measure $\mu_H=\mu(H,\varphi,\phi)$ as
\begin{align*}
    \mu_H(B) = \langle \varphi | P_H(B) | \phi \rangle, \quad \forall \mathcal  B(\R ).
\end{align*}

\section{Main results}
\label{sec:main}

In this section, we establish three theorems concerning the
extraction of the first spectral gap from the asymptotic
behavior of the commutator expectation along normalized
imaginary-time evolution.

Theorem~\ref{thm:main_2_gaps} considers
infinite-dimensional Hamiltonians whose ground-state and
first-excited energies are both isolated from the rest of
the spectrum. In this setting, under assumptions analogous
to those of \cite[Theorem~1]{Bondar2023}, the first
spectral gap can be recovered from the asymptotic decay.

Theorem~\ref{thm:main_1_gap} addresses the more delicate
case in which the ground-state energy is isolated, but
the first excited energy coincides with the threshold
of the continuous spectrum. Finally,
Theorem~\ref{thm:polynomial_density} provides a sufficient
condition on the spectral density near this threshold
and determines the resulting asymptotic behavior.

Before stating these results, we discuss the necessary
conditions on the initial state and observable.

In \cite[Theorem~1]{Bondar2023}, the finite-dimensional Hamiltonian is
represented by a discrete spectral decomposition,
with spectral projections $\Pi_0$ and $\Pi_1$
corresponding to the ground-state and first-excited
energies, respectively. The key assumption is
\[
    \operatorname{Im}
    \langle \psi,\Pi_0 O\Pi_1\psi\rangle
    \neq 0, \quad \Pi_0=P_H(\lambda_0), \quad \Pi_1=P_H(\lambda_1),
\]
where $\psi=\phi(0)$ is the initial state and $O$
is a self-adjoint observable.

This condition guarantees that the initial state has
nonzero projections onto both the ground-state and
first-excited eigenspaces, and that the observable
couples these components nontrivially. Consequently,
the contribution corresponding to the first spectral
gap appears with a nonzero coefficient in the
asymptotic expansion of the commutator expectation.

For Hamiltonians with continuous spectra, the situation
is more complicated. If the first excited energy
$\lambda_1$ is not an eigenvalue, then
\[
    P_H(\{\lambda_1\})=0,
\]
and the preceding condition cannot be imposed directly.
Instead, the excited-state contribution must be
described using the spectral measure of the Hamiltonian.

Let
\[
    \Pi_0=P_H(\{\lambda_0\}),
    \qquad Q=I-\Pi_0.
\]
At a minimum, the initial state must satisfy
\[
    \Pi_0\psi\neq0,
    \qquad Q\psi\neq0.
\]
Moreover, the observable must couple the ground-state
component to the excited spectral subspace, requiring
\[
    QO\Pi_0\psi\neq0.
\]

These conditions are necessary, but not sufficient,
for the recovery of the first spectral gap.
To see this, we consider the real signed spectral measure
\begin{align}
\label{eq:imaginary_part_measure}
    \nu_H(B)
    :=
    \operatorname{Im}
    \langle P_H(B)Q\psi,O\Pi_0\psi\rangle,
\end{align}
an object that becomes central during our analysis.
The measure vanishes identically whenever any of the
preceding nondegeneracy conditions fails.

However, even when all three conditions hold,
the measure $\nu_H$ may still vanish due to
cancellations of the imaginary parts. Furthermore,
its support need not extend to the bottom of the
continuous spectrum $\lambda_1$.

Thus, a nontrivial overlap between the ground-state
and excited spectral subspaces does not by itself
guarantee that the first spectral gap is detected.
What matters is the behavior of the associated
coupling spectral measure $\nu_H$ \eqref{eq:imaginary_part_measure} near $\lambda_1$.

Theorem~\ref{thm:main_1_gap} captures this dependence
through a Laplace-type integral associated with
$\nu_H$. Its asymptotic behavior determines whether
the logarithmic decay of the commutator expectation
reveals the first spectral gap.

Theorem~\ref{thm:polynomial_density} then provides
a more explicit sufficient condition: if the
coupling spectral measure has a nonvanishing
polynomial density near $\lambda_1$, the first
spectral gap can be recovered. Moreover, its
threshold behavior gives rise to an additional
logarithmic correction to the asymptotic decay,
which is absent in the corresponding
discrete-spectrum result.

\begin{theorem}
\label{thm:main_2_gaps}
[Asymptotics for infinite-dimensional Hamiltonian with two gaps.]
Let $H$ be a self-adjoint operator on a Hilbert space $\mathcal H$, bounded
from below, and suppose that
\[
    \sigma(H)
    \subset
    \{\lambda_0,\lambda_1\}\cup[\lambda_2,\infty),
    \qquad
    \lambda_0<\lambda_1<\lambda_2,
\]
equivalently the spectral decomposition reads
\[
H=\int_{\mathbb R}\lambda\,dP_\lambda
=
\lambda_0 P(\{\lambda_0\})
+
\lambda_1  P(\{\lambda_1\})
+
\int_{[\lambda_2,\infty)} \lambda\,dP_\lambda .
\]
Let us denote \begin{align*}
    \Pi_0 = P(\{\lambda_0\}), \quad \Pi_1 = P(\{\lambda_1\}).
\end{align*}
Let the initial state be
$\psi:=\phi(0)\in\mathcal H$.
For $\tau>0$, define the normalized imaginary-time evolution
\[
    \phi(\tau)
    :=
    \frac{e^{-\tau H}\psi}
         {\|e^{-\tau H}\psi\|}.
\]
Let $O\in B(\mathcal H)$ be self-adjoint. For $u\in D(H)$, define the
commutator form
\[
    \mathfrak c_O(u)
    :=
    \langle Hu,Ou\rangle-\langle Ou,Hu\rangle.
\]
If
\[
    \operatorname{Im}
    \langle \Pi_1\psi,O\Pi_0\psi\rangle
    \neq0,
\]
then
\[
    \mathfrak c_O(\phi(\tau))
    =
    \frac{
        2i(\lambda_1-\lambda_0)
        \operatorname{Im}
        \langle \Pi_1\psi,O\Pi_0\psi\rangle
    }{
        \|\Pi_0\psi\|^2
    }
    e^{-\tau(\lambda_1-\lambda_0)}
    +
    o\!\left(
        e^{-\tau(\lambda_1-\lambda_0)}
    \right).
\]
Consequently,
\[
    \ln
    \bigl|
        \mathfrak c_O(\phi(\tau))
    \bigr|
    =
    -\tau(\lambda_1-\lambda_0)
    +O(1),
    \qquad
    \tau\to\infty.
\]
\end{theorem}

\begin{remark}
Since $H$ is bounded from below, for every $\tau>0$ the operator
\[
    He^{-\tau H}
\]
is bounded. Consequently,
\[
    e^{-\tau H}\mathcal H\subset D(H),
\]
and hence $\phi(\tau)\in D(H)$ for every $\tau>0$. Therefore the
commutator form $\mathfrak c_O(\phi(\tau))$ is well defined for every
bounded operator $O$.

If, in addition,
\[
    O(D(H))\subset D(H),
\]
then the theorem may equivalently
be stated in terms of the operator commutator $[H,O]$
\[
    \mathfrak c_O(\phi(\tau))
    =
    \langle\phi(\tau),[H,O]\phi(\tau)\rangle.
\]
\end{remark}

\begin{proof}[Proof of Theorem \ref{thm:main_2_gaps}]

Since adding a real multiple of the identity does not change the
commutator form, we set
\[
    K:=H-\lambda_0 I.
\]
Moreover,
\[
    \frac{e^{-\tau H}\psi}{\|e^{-\tau H}\psi\|}
    =
    \frac{e^{-\tau K}\psi}{\|e^{-\tau K}\psi\|}.
\]
Set
\[
    \Delta:=\lambda_1-\lambda_0,
    \qquad
    \Gamma:=\lambda_2-\lambda_0,
\]
so that $0<\Delta<\Gamma$, and let
\[
    Q:=I-\Pi_0-\Pi_1.
\]
The spectral assumptions imply
\[
    K\Pi_0=0,
    \qquad
    K\Pi_1=\Delta\Pi_1,
    \qquad
    \sigma(K|_{Q\mathcal H})\subset[\Gamma,\infty).
\]

Write
\[
    \psi_0:=\Pi_0\psi,
    \qquad
    \psi_1:=\Pi_1\psi,
\]
and define
\[
    r_\tau:=e^{-\tau K}Q\psi.
\]
Then
\[
    u_\tau:=e^{-\tau K}\psi
    =
    \psi_0+e^{-\tau\Delta}\psi_1+r_\tau.
\]
By the spectral theorem,
\begin{align}
\label{eq:estimate_1}
    \|r_\tau\|
    \le
    e^{-\tau\Gamma}\|\psi\|.
\end{align}
Furthermore, although $K$ may be unbounded, the operator
$Ke^{-\tau K}Q$ is bounded for every $\tau>0$. Moreover, for any fixed
$\tau\geq\Gamma^{-1}$ 
\begin{align}
\label{eq:estimate_2}
    \|Ke^{-\tau K}Q\|
    &=
    \sup_{\lambda\in\sigma(K|_{Q\mathcal H})}
    \lambda e^{-\tau\lambda}
\le
    e^{-\tau\Gamma}
    \sup_{\lambda\ge\Gamma}
    \lambda e^{-\tau(\lambda-\Gamma)}
\le
        \Gamma
    e^{-\tau\Gamma}.
\end{align}
Hence
\begin{align}
\label{eq:estimate_3}
    \|Kr_\tau\|
    =
    O(e^{-\tau\Gamma}).
\end{align}

We now compute the commutator form. Since $O$ is self-adjoint,
\begin{align*}
    \mathfrak c_O(u_\tau)
    &=
    \langle Ku_\tau,Ou_\tau\rangle
    -
    \langle Ou_\tau,Ku_\tau\rangle
=
    2i\,
    \operatorname{Im}
    \langle Ku_\tau,Ou_\tau\rangle.
\end{align*}
Using
\[
    Ku_\tau
    =
    \Delta e^{-\tau\Delta}\psi_1
    +
    Kr_\tau,
\]
and
\[
    Ou_\tau
    =
    O\psi_0
    +
    e^{-\tau\Delta}O\psi_1
    +
    Or_\tau,
\]
we obtain
\begin{align*}
    \langle Ku_\tau,Ou_\tau\rangle
    &=
    \Delta e^{-\tau\Delta}
    \langle\psi_1,O\psi_0\rangle
    +
    \Delta e^{-2\tau\Delta}
    \langle\psi_1,O\psi_1\rangle
    +
    O(e^{-\tau\Gamma}).
\end{align*}
Since $O$ is self-adjoint,
\[
    \langle\psi_1,O\psi_1\rangle\in\mathbb R,
\]
and therefore the second term does not contribute to the imaginary
part. Consequently,
\[
    \mathfrak c_O(u_\tau)
    =
    2i\Delta\,
    \operatorname{Im}
    \langle\psi_1,O\psi_0\rangle
    e^{-\tau\Delta}
    \left( 1 + 
    O(e^{-\tau(\Gamma - \Delta)})
    \right).
\]

It remains to compute the normalization. By orthogonality of the spectral
subspaces,
\begin{align*}
    \|u_\tau\|^2
    &=
    \|\psi_0\|^2
    +
    e^{-2\tau\Delta}\|\psi_1\|^2
    +
    \|r_\tau\|^2
=
    \|\psi_0\|^2
    +
    O(e^{-2\tau\Delta}),
\end{align*}
from which
\begin{align*}
    \frac{1}{ \|u_\tau\|^2}
=
    \frac{1}{ \|\psi_0\|^2
   }
   \left( 1
   +
   O(e^{-2\tau\Delta})
   \right).
\end{align*}
Therefore
\begin{align*}
    \mathfrak c_O(\phi(\tau))
    &=
    \frac{\mathfrak c_O(u_\tau)}
         {\|u_\tau\|^2}
         \\
         &
         =
         \frac{ 2i\Delta\,
    \operatorname{Im}
    \langle\psi_1,O\psi_0\rangle
    e^{-\tau\Delta}}{||\psi_0||^2}
    \left( 1 + 
    O(e^{-\tau(\Gamma - \Delta)})
    \right)
     \left( 1
   +
   O(e^{-2\tau\Delta})
   \right), \quad \tau \to \infty,
   \\
   \ln \begin{vmatrix}\mathfrak c_O(\phi(\tau)) \end{vmatrix} &=
   - \tau \Delta + \ln 
   \begin{vmatrix}\frac{ 2\Delta\,
    \operatorname{Im}
    \langle\psi_1,O\psi_0\rangle}{||\psi_0||^2}
    \end{vmatrix}
    + o(1), \quad \tau \to \infty.
\end{align*}
Substituting
\[
    \Delta=\lambda_1-\lambda_0,
    \qquad
    \psi_j=\Pi_j\psi,
\]
gives the claimed asymptotic formula.

Finally, if
\[
    \operatorname{Im}
    \langle\Pi_1\psi,O\Pi_0\psi\rangle
    \neq0,
\]
the leading coefficient is nonzero, and hence
\[
    \ln
    \bigl|
        \mathfrak c_O(\phi(\tau))
    \bigr|
    =
    -\tau(\lambda_1-\lambda_0)+O(1).
\]
\end{proof}

\begin{theorem}
\label{thm:main_1_gap}[Asymptotics for infinite-dimensional Hamiltonian with one gap.]
Let \(H\) be a self-adjoint operator on a Hilbert space \(\mathcal H\),
bounded from below, and suppose that
\[
    \sigma(H)
    \subseteq
    \{\lambda_0\}\cup[\lambda_1,\lambda_2],
    \qquad
    -\infty<\lambda_0<\lambda_1<\lambda_2\leq+\infty, \qquad \lambda_0,\lambda_1 \in \sigma(H),
\]
equivalently the spectral decomposition reads
\[
H=\int_{\mathbb R}\lambda\,dP_H(\lambda)
=
\lambda_0 P(\{\lambda_0\})
+
\int_{[\lambda_1,\lambda_2]} \lambda\,dP_H(\lambda) .
\]
Let
\[
    \Pi_0 := P_H(\{\lambda_0\}),
    \qquad
    Q := I-\Pi_0,
    \qquad
    \Delta := \lambda_1-\lambda_0>0.
\]
Let the initial state \(\psi:=\phi(0)\in\mathcal H\) satisfy
\[
    \Pi_0\psi\neq 0, \quad (I -\Pi_0) \phi \not =0,
\]
and define the normalized imaginary-time evolved state by
\[
    \phi(\tau)
    :=
    \frac{e^{-\tau H}\psi}
    {\|e^{-\tau H}\psi\|},
    \qquad \tau>0.
\]
Let $O\in B(\mathcal H)$ be self-adjoint. For $u\in D(H)$, define the
commutator form
\[
    \mathfrak c_O(u)
    :=
    \langle Hu,Ou\rangle-\langle Ou,Hu\rangle.
\]
Define the finite real signed Borel measure
\[
    \nu_H(B)
    :=
    \operatorname{Im}
    \left\langle
        P_H(B)Q\psi,\,
        O\Pi_0\psi
    \right\rangle,
    \qquad
    B\in\mathcal B(\mathbb R).
\]
Assume that
\[
    F(\tau)
    :=
    \int_{[\lambda_1,\lambda_2]}
        (\lambda-\lambda_0)
        e^{-\tau(\lambda-\lambda_1)}
        \,d\nu_H(\lambda)
\]
is nonzero for all sufficiently large $\tau$, and that
\begin{align}
\label{eq:Laplace_integral_condition}
    \lim_{\tau\to\infty}
    \frac{1}{\tau}\ln|F(\tau)|
    =0.
\end{align}
Then
\begin{align}
\label{eq:main_1_gap_3}
     \mathfrak c_O(\phi(\tau))
    =
    \frac{2i}{\|\Pi_0\phi_0\|^2}
    e^{-\tau\Delta}
    F(\tau)
    \left(
        1+o(1)
    \right),
    \qquad
    \tau\to\infty.
\end{align}
Consequently,
\begin{align}
\label{eq:main_1_gap_1}
        \ln
    \bigl|
        \mathfrak c_O(\phi(\tau))
    \bigr|
   & =
    -\tau(\lambda_1-\lambda_0)
    +
    \ln|F(\tau)|
    +O(1),
    \qquad
    \tau\to\infty,
    \\ \label{eq:main_1_gap_2}
    \lim_{\tau\to\infty}
    \frac{ \ln
    \bigl|
        \mathfrak c_O(\phi(\tau))
    \bigr|}{\tau}& = -(\lambda_1-\lambda_0)  
    .
\end{align}
\end{theorem}
\begin{proof}
The main steps of the proof are the same as those of the proof of Theorem \ref{thm:main_2_gaps}.
The crucial difference is that now, the second smallest point of spectrum $\lambda_1$ is not in general associated with a projection $\Pi_1$ onto its eigenspace.
    Since adding a scalar multiple of the identity does not change the
commutator form, we again set
\[
    K:=H-\lambda_0 I, \quad
    \Delta:=\lambda_1-\lambda_0, \quad
    Q:=I-\Pi_0.
\]
Let $P_K$ denote the projection-valued spectral measure of $K$.
Since
    $K=H-\lambda_0 I,$
the spectral measures of $H$ and $K$ are related by translation:
\[
    P_K(B)
    =
    P_H(B+\lambda_0),
    \qquad
    B\in\mathcal B(\mathbb R).
\]
Define the corresponding signed coupling measure for $K$ by
\[
    \nu_K(B)
    :=
    \operatorname{Im}
    \left\langle
        P_K(B)Q\psi,\,
        O\psi_0
    \right\rangle.
\]
Then
\begin{align}
\label{eq:spectral_shift}
    \nu_K(B)
    =
    \nu_H(B+\lambda_0).
\end{align}
The spectral assumptions imply
\[
    K\Pi_0=0,
    \qquad
    \sigma(K|_{Q\mathcal H})\subset[\Delta,\infty).
\]

Write
\[
    \psi_0:=\Pi_0\psi,
\]
and define
\[
    r_\tau:=e^{-\tau K}Q\psi.
\]
Then
\[
    u_\tau:=e^{-\tau K}\psi
    =
    \psi_0+r_\tau, \quad K u_\tau  = K r_\tau.
\]
Repeating the same steps as in \eqref{eq:estimate_1}-\eqref{eq:estimate_3}, we obtain that 
\begin{align}
\label{eq:estimate_4}
    \begin{Vmatrix}
        r_\tau
    \end{Vmatrix}
    \leq e^{-\tau \Delta} \begin{Vmatrix}
        \psi
    \end{Vmatrix}, \quad 
    \begin{Vmatrix}
        K e^{-\tau K} Q 
    \end{Vmatrix}
    \leq \Delta e^{-\tau \Delta}, \quad \begin{Vmatrix}
        K r_\tau
    \end{Vmatrix} = O(e^{-\tau \Delta}).
\end{align}

The commutator form is 
\begin{align*}
    \mathfrak c_O(u_\tau)
    &=
    2i\,
    \operatorname{Im}
    \langle Ku_\tau,Ou_\tau\rangle
    \\
    &=2i \operatorname{Im}
    \langle K r_\tau, O (\psi_0 + r_\tau) \rangle
    \\
    &=
    2i \operatorname{Im}
    \left(
    \langle
    Ke^{-\tau K}Q \psi, O \psi_0 \rangle
    +
     \langle
    K r_\tau, O r_\tau \rangle
    \right)
    .
\end{align*}
We inspect the two terms separately. For the second term, due to \eqref{eq:estimate_4}, we have 
\begin{align*}
    \begin{vmatrix}
        \langle
    K r_\tau, O r_\tau \rangle
    \end{vmatrix}
    = O(e^{-2\tau\Delta}).
\end{align*}
Now, we analyze the first term.
Using the Spectral Theorem from Section \ref{sec:spectral_theorem}, we assign the spectral complex-valued measure $(K,\varphi, \phi) \mapsto \mu_K =\mu({K,\varphi,\phi})$ for $\varphi = Q\psi, \ \phi = O \psi_0$.
Let $\nu_K$ be the imaginary part of the measure $\mu_K$.
\begin{align} 
\nonumber
\operatorname{Im}\left(
    \langle
    Ke^{-\tau K}Q \psi, O \psi_0 \rangle \right) =
    e^{-\tau \Delta}
    \operatorname{Im}\left(
    \int_{\Delta}^{\infty}s e^{-\tau (s-\Delta)} d \mu_K(s)\right)
    \\ \label{eq:integral_term}
    =
    e^{-\tau \Delta}
    \int_{\Delta}^{\infty}s e^{-\tau (s-\Delta)} d \nu_K(s)
    =
    e^{-\tau \Delta}
    \int_{\lambda_1}^{\infty} (\lambda-\lambda_0)e^{-\tau(\lambda-\lambda_1)} d \nu_H(\lambda)
    = e^{-\tau\Delta} F(\tau),
\end{align}
where we used substitution $s=\lambda-\lambda_0$ and the fact that $\tilde \nu$ and $\nu$ are related by \eqref{eq:spectral_shift}.
This leads to 
\begin{align}
\label{eq:desired_asymptotics}
        \mathfrak c_O(u_\tau)
    = 2i
    e^{-\tau \Delta }
 F(\tau) 
    \left(
    1 + O\left(
    \frac{e^{-\tau\Delta}}{F(\tau)}
    \right)
    \right)
    \\
    \ln \begin{vmatrix}
        \mathfrak c_O(u_\tau)
    \end{vmatrix}
    =
    -\tau \Delta + \ln
    \begin{vmatrix}
        F(\tau)
    \end{vmatrix}+\ln2
    +
     O\left(
    \frac{e^{-\tau\Delta}}{F(\tau)}
    \right).
\end{align}
Dividing  by $\tau$, we obtain
\[
    \frac{1}{\tau}
    \ln
    \left|
        \mathfrak c_O(u_\tau)
    \right|
    =
    -\Delta
    +
    \frac{\ln|F(\tau)|}{\tau}
    +
    \frac{\ln2}{\tau}
    +
    O\left(
    \frac{e^{-\tau\Delta}}{\tau F(\tau)}\right).
\]

Let us show that
\begin{align}
\label{eq:Landau_relationship}
O\left(
    \frac{e^{-\tau\Delta}}{ F(\tau)}
    \right) = o(1), \quad 
     O\left(
    \frac{e^{-\tau\Delta}}{\tau F(\tau)}
    \right) = o(1), \quad \tau \to \infty.
\end{align}
By assumption,
\[
    \lim_{\tau\to\infty}
    \frac{\ln |F(\tau)|}{\tau}
    =0.
\]
Hence, for every $\varepsilon>0$, there exists $\tau_\varepsilon>0$
such that for all $\tau\geq\tau_\varepsilon$,
\[
    \frac{\ln |F(\tau)|}{\tau}>-\varepsilon,
\]
or equivalently,
\[
    |F(\tau)|>e^{-\varepsilon\tau}.
\]
Choose $0<\varepsilon<\Delta$. Then
\[
    \frac{e^{-\tau\Delta}}{\tau |F(\tau)|}
    \leq 
    \frac{e^{-\tau\Delta}}{ |F(\tau)|}
    \leq \frac{    e^{-(\Delta-\varepsilon)\tau}}{\tau}
    \longrightarrow 0,
    \qquad
    \tau\to\infty.
\]
Therefore, all of \eqref{eq:Landau_relationship} holds and
\[
    \lim_{\tau\to\infty}
    \frac{1}{\tau}
    \ln
    \left|
        \mathfrak c_O(u_\tau)
    \right|
    =
    -\Delta.
\]

It remains to take into account the normalization.
Since
\[
    u_\tau
    =
    \psi_0+r_\tau,
\]
with
\[
    \psi_0\in\Pi_0\mathcal H,
    \qquad
    r_\tau\in Q\mathcal H,
\]
we have
\[
    \langle \psi_0,r_\tau\rangle=0.
\]
Therefore,
\[
    \|u_\tau\|^2
    =
    \|\psi_0\|^2+\|r_\tau\|^2
    =
    \|\psi_0\|^2\left( 1 
    +
    O(e^{-2\tau\Delta}) \right).
\]
Since $\psi_0=\Pi_0\psi\neq0$,
\[
   \frac{1}{\|u_\tau\|^2}
   =
   \frac{1}{ \|\psi_0\|^2}
   \left( 1 
    +
    O(e^{-2\tau\Delta}) \right)
\]

Recalling that
\[
    \phi(\tau)
    =
    \frac{u_\tau}{\|u_\tau\|},
\]
we obtain
\[
    \mathfrak c_O(\phi(\tau))
     = \frac{\mathfrak c_O(u_\tau)}{\begin{Vmatrix}
        u_\tau
    \end{Vmatrix}^2}
    =
    \frac{2i}{\|\Pi_0\psi\|^2}
    e^{-\tau \Delta }
 F(\tau) 
    \left(
    1 + O\left(
    \frac{e^{-\tau\Delta}}{F(\tau)}
    \right)
    \right)
.
\]
Taking logarithm of the absolute value yields the desired asymptotics 
\eqref{eq:main_1_gap_1}
and dividing by $\tau$ reveals the spectral gap in the limit \eqref{eq:main_1_gap_2}.
\end{proof}

\begin{theorem}
\label{thm:polynomial_density}
[Asymptotics for Hamiltonians with polynomial spectral measure.]
    Let $\mathcal H,H,\psi,O$ satisfy the structural assumptions of
Theorem~\ref{thm:main_1_gap}, but replace the assumptions on $F(\tau)$
by the following threshold-density assumption.
    Suppose that for some \(\varepsilon>0\) the restriction of
\(\nu_H\) to \([\lambda_1,\lambda_1+\varepsilon]\) is absolutely continuous,
\[
    d\nu_H(\lambda)=\rho(\lambda)\,d\lambda,
\]
and that there exist \(c\in\mathbb R\setminus\{0\}\) and
\(\alpha>-1\) such that
\[
    \rho(\lambda_1+x)
    =
    c\,x^\alpha + o(x^\alpha),
    \qquad
    x\downarrow 0.
\]
Then
\[
    F(\tau)
    =
    \tau^{-(\alpha+1)}
    \left(
    c\Delta\Gamma(\alpha+1)
    +o(1)
    \right),
    \qquad
    \tau\to\infty,
\]
and the condition \eqref{eq:Laplace_integral_condition} from Theorem \ref{thm:main_1_gap} is satisfied.
Consequently the commutator term from \eqref{eq:main_1_gap_3} becomes
\[
    \mathfrak c_O(\phi(\tau))
    =
    e^{-\tau\Delta}
    \tau^{-(\alpha+1)}
    \left(
    \frac{
        2ic\Delta\Gamma(\alpha+1)
    }{
        \|\Pi_0\psi\|^2
    } + o(1)
    \right).
\]
Hence
\[
    \ln
    \left|
        \mathfrak c_O(\phi(\tau))
    \right|
    =
    -(\lambda_1-\lambda_0)\tau
    -(\alpha+1)\ln\tau
    +O(1).
\]
\end{theorem}
\begin{proof}
From \eqref{eq:integral_term} in the proof of
Theorem~\ref{thm:main_1_gap},
\[
    F(\tau)
    =
    \int_{[\lambda_1,\lambda_2]}
    (\lambda-\lambda_0)
    e^{-\tau(\lambda-\lambda_1)}
    \,d\nu_H(\lambda).
\]
We split
\[
    F(\tau)=F_1(\tau)+F_2(\tau),
\]
where
\begin{align*}
    F_1(\tau)
    &:=
    \int_{\lambda_1}^{\lambda_1+\varepsilon}
    (\lambda-\lambda_0)
    e^{-\tau(\lambda-\lambda_1)}
    \rho(\lambda)\,d\lambda,
    \\
    F_2(\tau)
    &:=
    \int_{[\lambda_1+\varepsilon,\lambda_2]}
    (\lambda-\lambda_0)
    e^{-\tau(\lambda-\lambda_1)}
    \,d\nu_H(\lambda).
\end{align*}

We first estimate $F_2$. Since $\nu_H$ is a finite signed measure,
its total variation $|\nu_H|$ is finite. For sufficiently large
$\tau$, the function
\[
    \lambda\longmapsto
    (\lambda-\lambda_0)
    e^{-\tau(\lambda-\lambda_1)}
\]
is decreasing on $[\lambda_1+\varepsilon,\infty)$. Hence
\begin{align*}
    |F_2(\tau)|
    &\leq
    \int_{[\lambda_1+\varepsilon,\lambda_2]}
    (\lambda-\lambda_0)
    e^{-\tau(\lambda-\lambda_1)}
    \,d|\nu_H|(\lambda)
    \\
    &\leq
    (\Delta+\varepsilon)
    e^{-\tau\varepsilon}
    |\nu_H|(\mathbb R).
\end{align*}
Therefore,
\[
    F_2(\tau)=O(e^{-\tau\varepsilon}).
\]

We now consider $F_1$. Setting
\[
    x:=\lambda-\lambda_1,
\]
we obtain
\[
    F_1(\tau)
    =
    \int_0^\varepsilon
    (\Delta+x)e^{-\tau x}
    \rho(\lambda_1+x)\,dx.
\]
By assumption,
\[
    \rho(\lambda_1+x)
    =
    c x^\alpha+o(x^\alpha)
    =
    x^\alpha\bigl(c+\eta(x)\bigr), \quad \eta(x)\longrightarrow0,
    \qquad x\downarrow0.
\]
Thus
\[
    F_1(\tau)
    =
    \int_0^\varepsilon
    (\Delta+x)e^{-\tau x}
    x^\alpha
    \bigl(c+\eta(x)\bigr)\,dx.
\]
We now make the substitution
$y=\tau x$.
Then
\begin{align*}
    F_1(\tau)
    &=
    \tau^{-(\alpha+1)}
    \int_0^{\tau\varepsilon}
    \left(
        \Delta+\frac{y}{\tau}
    \right)
    e^{-y}y^\alpha
    \left(
        c+\eta\left(\frac{y}{\tau}\right)
    \right)
    dy.
\end{align*}

Using the Lebesgue's Dominated Convergence Theorem, we show that
\[
    \int_0^{\tau\varepsilon}
    \left(
        \Delta+\frac{y}{\tau}
    \right)
    e^{-y}y^\alpha
    \left(
        c+\eta\left(\frac{y}{\tau}\right)
    \right)
    dy
    \longrightarrow
    c\Delta
    \int_0^\infty e^{-y}y^\alpha\,dy.
\]
Indeed, for each fixed $y>0$,
\[
    \left(
        \Delta+\frac{y}{\tau}
    \right)
    \left(
        c+\eta\left(\frac{y}{\tau}\right)
    \right)
    \longrightarrow
    c\Delta.
\]
Moreover, after decreasing $\varepsilon>0$ if necessary,
$c+\eta(x)$ is bounded on $(0,\varepsilon]$. Hence the integrand,
extended by zero for $y>\tau\varepsilon$, is dominated by 
\[
   C y^{\alpha}  e^{-y} \in L^1((0,+\infty)),
\]
for some constant $C >0$. Since $\alpha>-1$, this function is integrable.
Therefore, 
\[
    F_1(\tau)
    \sim
    c\Delta\Gamma(\alpha+1)
    \tau^{-(\alpha+1)}.
\]

Since
\[
    F_2(\tau)=O(e^{-\tau\varepsilon})
\]
and
\[
    e^{-\tau\varepsilon}
    =
    o\left(\tau^{-(\alpha+1)}\right),
\]
we conclude that
\[
    F(\tau)
    =
    \tau^{-(\alpha+1)}
    \left(
    c\Delta\Gamma(\alpha+1)
    +o(1)
    \right)
    .
\]

In particular, since
\[
    c\Delta\Gamma(\alpha+1)\neq0,
\]
the function $F(\tau)$ is nonzero for all sufficiently large $\tau$,
and
\begin{align*}
    \ln|F(\tau)|
    &=
    -(\alpha+1)\ln\tau
    +
    \ln\left|
        c\Delta\Gamma(\alpha+1)
    \right|
    +
    o(1).
\end{align*}
Consequently,
\[
    \lim_{\tau\to\infty}
    \frac{\ln|F(\tau)|}{\tau}
    =0.
\]
Thus the assumptions on $F$ in
Theorem~\ref{thm:main_1_gap} are satisfied.

Applying Theorem~\ref{thm:main_1_gap}, we obtain
\[
    \mathfrak c_O(\phi(\tau))
    =
     e^{-\tau\Delta}
    \tau^{-(\alpha+1)}
    \left(
    \frac{
        2ic\Delta\Gamma(\alpha+1)
    }{
        \|\Pi_0\psi\|^2
    } + o(1)
    \right)
   .
\]
Therefore,
\begin{align*}
    \ln
    \left|
        \mathfrak c_O(\phi(\tau))
    \right|
    &=
    -\tau\Delta
    -(\alpha+1)\ln\tau
    +O(1)
    \\
    &=
    -(\lambda_1-\lambda_0)\tau
    -(\alpha+1)\ln\tau
    +O(1),
\end{align*}
which proves the claim.
\end{proof}

\newpage

\section{Application to the one-center point interaction}
\label{sec:point_interaction}

As a concrete application of Theorem~\ref{thm:polynomial_density},
we consider the attractive one-center point interaction in three
dimensions.
Formally, this Hamiltonian is 
\begin{align*}H_y=
    -\Delta + \delta_y, \quad \text{in } L^2(\R^3). 
\end{align*}
However, following \cite[Theorem~I.1.1 and Eq.~(1.1.15)]{AlbeverioEtAl}, correct functional analysis introduces the operator as
\begin{align*}
    -\Delta_{C_0^\infty(\R^3\setminus\{y\})}, \quad y \in \R^3.
\end{align*}

\cite{AlbeverioEtAl} uses von Neumann self-adjoint extension theory, where they show that both of the defect indices are precisely one.
Therefore, all the self-adjoint of extensions of $H_y$ are given by a one-parameter family.

Decomposing the Hilbert space using spherical coordinates, one obtains
\begin{align*}
    L^2(\R^3)= L^2((0,\infty),r^2dr) \otimes L^2(S^2),
\end{align*}
where $S^2$ is the unit sphere in $\R^3$.
According to \cite{AlbeverioEtAl}, studying the one-parametric self-adjoint extensions of $H_y$ is then reduced to only the radial part Laplacian with Robin boundary condition
\begin{align}
\label{eq:differential equation}
    H_\beta
    =
    -\frac{d^2}{dr^2}, \qquad \text{ in } L^2((0,+\infty))
    \\
 D(H_\beta)=
\begin{Bmatrix}
    f \in L^2((0,\infty))| f,f'\in \operatorname{AC}_{loc}((0,\infty)); \beta f(0+)=f'(0+); f'' \in L^2((0,+\infty))
\end{Bmatrix}
\end{align}
see \cite[Eq.~(1.1.12)]{AlbeverioEtAl}.

We consider the attractive case $\beta<0$. By
\cite[Theorem~I.1.4]{AlbeverioEtAl},
\begin{align}
\label{eq:spectrum_point_interaction}
    \sigma(h_\beta)
    =
    \{-\beta^2\}\cup[0,\infty),
\end{align}
where $-\beta^2$ is a simple eigenvalue and the spectrum on
$[0,\infty)$ is purely absolutely continuous. Hence, in the notation
of Theorem~\ref{thm:polynomial_density},
\[
    \lambda_0=-\beta^2,
    \qquad
    \lambda_1=0,
    \qquad
    \Delta=\lambda_1-\lambda_0=\beta^2.
\]

Now, we analyze the spectral measures for a fixed $\beta$, therefore we write shortly $H$ for $H_\beta$. 
This means for $f\in L^2((0,+\infty))$, $g \in \operatorname{Dom}(H) $, to find $\mu = \mu(H, f,g)$
\begin{align*}
\langle f, H g \rangle 
=
\int_{\sigma(H)}
\lambda d \mu_{H,f,g}(\lambda), 
\\
\mu_{H,f,g}(B) = \langle f, P_{H}(B)g\rangle, \quad B \in \mathcal{B}(\R).
\end{align*}

For $\lambda=k^2>0$, the generalized  eigenfunctions are explicitly
known; see
\cite[Section~I.1.4, Eqs.~(I.1.4.1)--(I.1.4.5)]{AlbeverioEtAl}.
For a fixed $\beta<0$, a convenient real boundary-normalized solution to \eqref{eq:differential equation} is
\[
    \vartheta_k(r)
    =
    \cos(kr)+\frac{\beta}{k}\sin(kr).
\]
Since $\vartheta_k''(r)=-k^2\vartheta_k(r)$ and we know the spectrum of $H$, we expect a representation of a form 
\begin{align}
\label{eq:measure_form}
    \langle f, H g \rangle  = - \beta^2 a_{f,g} + \int_0^\infty k^2 d \nu_{f,g}(k).
\end{align}

To find the measure $\nu_{H,f,g}$ from \eqref{eq:measure_form} explicitly, we use the generalized Fourier expansion of the half-line Robin
Laplacian  established in
\cite[Theorem~2.1(b), Eq. (2.2) and Eqs.~(2.7)--(2.9)]{Bouzeffour2026}.
Denoting
\[
    \mathcal C_\beta f(k)
    =
    \int_0^\infty f(r)
    \left(
        \cos(kr)+\frac{\beta}{k}\sin(kr)
    \right)\,dr,
\]
the corresponding Plancherel identity and diagonalization
formula yield
\begin{align*}
    \langle f,Hg\rangle
    &=
    -\beta^2
    \overline{\langle\phi_\beta,f\rangle}
    \langle\phi_\beta,g\rangle +
    \int_0^\infty k^2
    \overline{\mathcal C_\beta f(k)}
    \mathcal C_\beta g(k)\,
    d\rho_\beta(k),
    \\
     d\rho_\beta(k)
    &=
    \frac{2}{\pi}
    \frac{k^2}{k^2+\beta^2}\,dk.
    \\
    \phi_\beta(r) &= \sqrt{-2\beta} e^{\beta r},
\end{align*}
from which we identify the constant $a_{f,g}$ as 
\begin{align*}
    a_{f,g} = \overline{\langle\phi_\beta,f\rangle}
    \langle\phi_\beta,g\rangle.
\end{align*}
The continuous spectral
measure $\nu_{f,g}$  is
\[
    d\nu_{f,g}(k)
    =
    \overline{\mathcal C_\beta f(k)}
    \mathcal C_\beta g(k)\,d\rho_\beta(k).
\]
Consequently,
\[
    \frac{d\nu_{f,g}}{dk}(k)
    =
    \frac{2}{\pi}
    \frac{k^2}{k^2+\beta^2}
    \overline{\mathcal C_\beta f(k)}
    \mathcal C_\beta g(k), \quad k>0.
\]

We now investigate the behavior of the continuous spectral
density as $k\to 0^+$.
Assume that $f,g$ satisfy
\[
    \int_0^\infty
    (1+r)(|f(r)|+|g(r)|)\,dr<\infty.
\]
Since
\[
    \lim_{k\to0^+}
    \left(
        \cos(kr)+\frac{\beta}{k}\sin(kr)
    \right)
    =
    1+\beta r
\]
and
\[
    \left|
        \cos(kr)+\frac{\beta}{k}\sin(kr)
    \right|
    \leq 1+|\beta|r,
\]
the dominated convergence theorem yields
\[
    \lim_{k\to0^+}\mathcal C_\beta f(k)
    =
    \int_0^\infty(1+\beta r)f(r)\,dr
    =:A_\beta(f).
\]
Consequently,
\[
    \mathcal C_\beta f(k)
    =
    A_\beta(f)+o(1),
    \qquad k\to0^+,
\]
and analogously for $g$.

Substituting these expansions into the spectral density,
we obtain
\begin{align*}
    \frac{d\nu_{f,g}}{dk}(k)
    &=
    \frac{2}{\pi}
    \frac{k^2}{k^2+\beta^2}
    \left(
        \overline{A_\beta(f)}A_\beta(g)
        +o(1)
    \right)\\
    &=
    \frac{
        2\overline{A_\beta(f)}A_\beta(g)
    }{\pi\beta^2}k^2
    +o(k^2),
    \qquad k\to0^+,
\end{align*}
where we used $\beta<0$ and, in particular,
$\beta^2>0$.

We have thus established the asymptotic behavior
of the continuous spectral density in the momentum
variable $k$.

To express the result in terms of the original
spectral parameter $\lambda$ by making the substitution
\[
    \lambda=k^2,
    \qquad
    dk=\frac{1}{2\sqrt{\lambda}}\,d\lambda,
\]
we obtain
\begin{align*}
    \frac{d\mu_{H,f,g}^{\mathrm{ac}}}{d\lambda}
        (\lambda)
    &=
    \frac{1}{2\sqrt{\lambda}}
    \frac{d\nu_{f,g}}{dk}
        (\sqrt{\lambda})\\
    &=
    \frac{\sqrt{\lambda}}
         {\pi(\lambda+\beta^2)}
    \overline{
        \mathcal C_\beta f(\sqrt{\lambda})
    }
    \mathcal C_\beta g(\sqrt{\lambda}).
\end{align*}
Using the previously established asymptotics,
we conclude that
\begin{equation}
\label{eq:point_interaction_threshold_density}
\frac{d\mu_{H,f,g}^{\mathrm{ac}}}{d\lambda}
        (\lambda)
    =
    \frac{
        \overline{A_\beta(f)}A_\beta(g)
    }{\pi\beta^2}
    \lambda^{1/2}
    +o(\lambda^{1/2}),
    \qquad \lambda\to0^+.
\end{equation}

Now, we make the connection to satisfying the assumptions of Theorem \ref{thm:polynomial_density}.
Let $\psi \in D(H)$ be an initial state and $O$ and observable.
In particular, 
provided that
$A_\beta(f)A_\beta(g)\neq0$ with the choice 
\begin{align*}
    f:= Q \psi, \quad  g:= O \Pi_0 \psi,
\end{align*}
the leading
coefficient is nonzero, and the exponent from Theorem \ref{thm:polynomial_density} is
\[
\alpha=\frac12.
\]

\begin{remark}
The nonvanishing condition
\[
    A_\beta(f)A_\beta(g)\neq 0
\]
ensures that the vectors $f$ and $g$ have nontrivial
spectral components arbitrarily close to the threshold
$\lambda_1=0$.

Indeed, by the spectral representation, for
$\varepsilon>0$,
\begin{align*}
    \|P_H([0,\varepsilon))f\|^2
    &=
    \int_0^{\sqrt{\varepsilon}}
    \frac{2}{\pi}
    \frac{k^2}{k^2+\beta^2}
    |\mathcal C_\beta f(k)|^2\,dk\\
    &=
    \frac{2|A_\beta(f)|^2}{3\pi\beta^2}
    \varepsilon^{3/2}
    +o(\varepsilon^{3/2}).
\end{align*}
Consequently, if $A_\beta(f)\neq0$, then
\[
    P_H([\lambda_1,\lambda_1+\varepsilon))f
    \neq0
    \qquad\text{for every }\varepsilon>0.
\]
The same conclusion holds for $g$ if
$A_\beta(g)\neq0$.

Thus, the nonvanishing of the leading threshold
coefficient implies that both $f$ and $g$ have
nonzero spectral projections in every neighborhood
of $\lambda_1$.

The converse need not hold: nonzero spectral
projections arbitrarily close to $\lambda_1$
do not necessarily imply that
$A_\beta(f)A_\beta(g)\neq0$.
\end{remark}

\subsection{Numerical example}
\label{sec:numerics}

We have established that the Robin Laplacian model satisfies
the assumptions of Theorem \ref{thm:polynomial_density}
with the exponent $\alpha=0.5$.
Therefore, the theoretical prediction for the asymptotic
behavior of the commutator is
\[
    \ln\left|\mathfrak c_O(\phi(\tau))\right|
    =
    -\kappa^2\tau
    -\frac32\ln\tau
    +C+o(1),
    \qquad \tau\to\infty.
\]
Thus, a fit of the form
\begin{align}
\label{eq:fit_1}
    y(\tau)=y(\tau,a_0,a_1,a_2)
    =
    a_1\tau+a_2\ln\tau+a_0
\end{align}
should yield
\[
    a_1=-\kappa^2,
    \qquad
    a_2=-\frac32.
\]

We consider the Robin Laplacian
\[
    H=-\frac{d^2}{dx^2},
    \qquad
    u'(0)=-\kappa u(0),
    \qquad \kappa=0.5,
\]
whose spectral gap is $\Delta=\kappa^2=0.25$.
For the numerical approximation, we truncate the half-line
to $[0,L]$, where $L=20$, impose a Dirichlet boundary condition
at $x=L$, and discretize the Hamiltonian using piecewise-linear
finite elements with a lumped mass matrix and $N=1000$
spatial unknowns.

We randomly generate an initial state $h$ and two real-valued
functions $f,g$ from the family
\[
    u(x)=e^{-\gamma x^2}
    \left(
        1-\kappa x+\sum_{j=1}^{2}c_jx^{2j}
    \right),
    \qquad c_j\sim\mathcal N(0,1),
\]
with $\gamma=0.15$ and random seed $2026$.
For the initial state $h$, we allow complex Gaussian
coefficients. By construction, all generated functions
satisfy the Robin boundary condition.
After normalization and orthogonalization of $f$ and $g$,
we define the rank-two self-adjoint observable
\[
    O=i\left(
        |f\rangle\langle g|-|g\rangle\langle f|
    \right).
\]
We then compute the normalized imaginary-time evolution
\[
    \phi(\tau)=
    \frac{e^{-\tau H}h}{\|e^{-\tau H}h\|}
\]
and evaluate the commutator
$\mathfrak c_O(\phi(\tau))$ numerically.
The evolution is approximated using sparse matrix
exponentials, without diagonalizing the Hamiltonian
or supplying its spectral gap.
We use $1001$ equally spaced time points on $[0,60]$.

We compare three fitting approaches.
First, we fit $y(\tau)$ to
$\ln|\mathfrak c_O(\phi(\tau))|$.
Second, we fit $y(\tau)/\tau$ to
$\ln|\mathfrak c_O(\phi(\tau))|/\tau$.
Third, we fit
\[
    y'(\tau)=a_1+\frac{a_2}{\tau}
\]
to
\[
    \frac{d}{d\tau}
    \ln|\mathfrak c_O(\phi(\tau))|,
\]
where the derivative is approximated by finite differences.
The fitting parameters are determined by minimizing the
root mean square error (RMSE) on the interval
$\tau\in[25,35]$.
The first two approaches fit all three coefficients,
including the constant term $a_0$, while the third
fits only $a_1$ and $a_2$.


\begin{figure}[ht]
    \centering
    \includegraphics[width=0.9\textwidth]
        {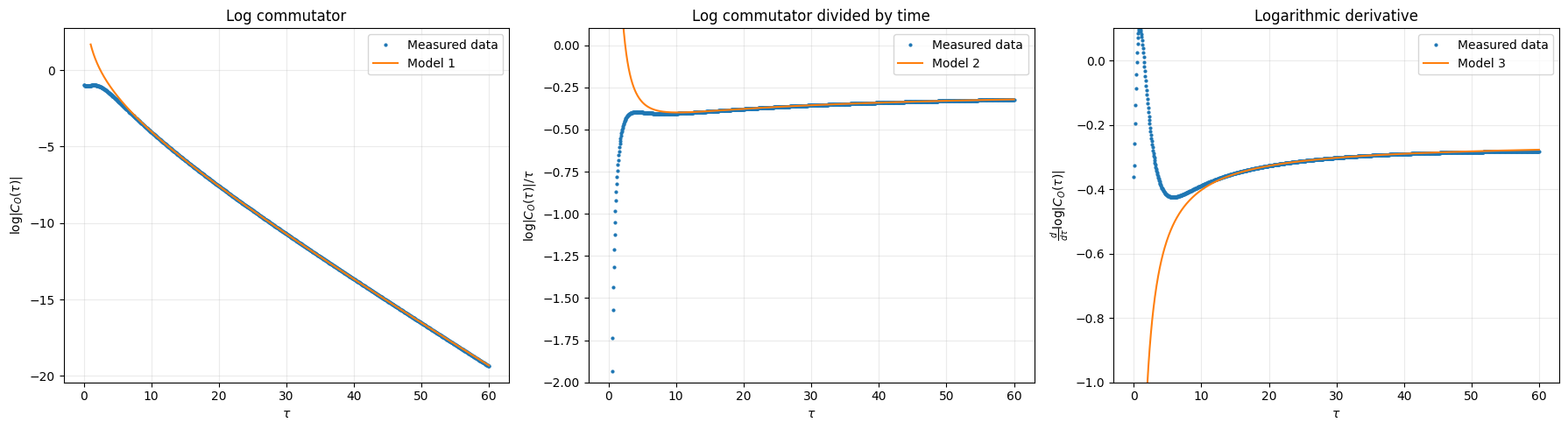}
    \caption{
        Imaginary-time decay of the commutator form for
        the Robin Laplacian.
        The asymptotic prediction is
        $-\kappa^2\tau-\frac32\ln\tau+O(1)$.
    }
    \label{fig:point_interaction_decay}
\end{figure}

\begin{table}[ht]
\centering
\begin{tabular}{lccc}
\hline
\textbf{Method} &
\textbf{$\Delta$} &
\textbf{$\alpha+1$} &
\textbf{Constant term} \\
\hline
Fit $y(\tau)$
& $0.252117 \pm 0.000015$
& $1.503846 \pm 0.000445$
& $1.969424 \pm 0.001063$ \\

Fit $y(\tau)/\tau$
& $0.252087 \pm 0.000014$
& $1.504756 \pm 0.000423$
& $1.970832 \pm 0.000530$ \\

Fit $y'(\tau)$
& $0.252163 \pm 0.000023$
& $1.502856 \pm 0.000674$
& --- \\
\hline
Theory
& $0.25$
& $1.5$
& --- \\
\hline
\end{tabular}
\caption{
    Numerical estimates of the spectral gap $\Delta$
    and threshold exponent $\alpha+1$ obtained using
    three fitting procedures.
    The fitting interval is $\tau\in[25,35]$.
}
\label{tab:numerical_fits}
\end{table}

The numerical estimates are in good agreement with
the theoretical values $\Delta=0.25$ and
$\alpha+1=1.5$.
All three fitting approaches yield similar results,
with relative deviations of approximately $0.85\%$
for the spectral gap and below $0.4\%$ for the
threshold exponent.
The reported uncertainties are formal least-squares
fit uncertainties; they do not account for systematic
errors due to spatial discretization, finite-domain
truncation, or pre-asymptotic corrections.

\section{Outlook}

Our results provide a novel computational tool for quantum simulations: in infinite-dimensional systems, the imaginary-time decay of the commutator expectation carries more information than the spectral gap alone. The coefficient $-(\alpha+1)$ of the logarithmic correction in Theorem~\ref{thm:polynomial_density} encodes the threshold exponent $\alpha$ of the coupling spectral measure. Fitting the ansatz $a_1\tau+a_2\ln\tau+a_0$ to data that existing ground-state algorithms already produce therefore yields both $\Delta=-a_1$ and $\alpha=-a_2-1$, with minimal changes to existing codes and at the computational cost of a ground-state calculation.

This matters because $\alpha$ is a fingerprint of the physical nature of dynamics above the gap. For a generic state--observable coupling, $\alpha$ is inherited from the density of states at the threshold. Bosonic excitations, such as phonons in three dimensions, have $\alpha=2$ (see, e.g., Eq.~(2.5) in Ref.~\cite{Simon2013}), which gives a correction $-3\ln\tau$. Free fermions have $\alpha=\tfrac12$ (see, e.g., Eq.~(4.11) in Ref.~\cite{Simon2013}), which gives a correction $-\tfrac32\ln\tau$. More generally, $\alpha=\tfrac{d}{2}-1$ for a regular quadratic band edge in $d$ dimensions, and $\alpha=l+\tfrac12$ in the radial channel with angular momentum $l$ (Table~\ref{tab:numerical_fits}). The logarithmic correction therefore distinguishes dynamics, the effective dimensionality, and the orbital angular momentum of the excitations.

Natural next steps include threshold behaviors beyond pure power laws, such as the logarithmic densities that arise in two dimensions.

\bibliography{sn-bibliography}

\end{document}